\documentstyle[12pt]{article}
\newcommand{\um}{~$\mu$m}
\newcommand{\afifty}{$\alpha_{1950.0}$}
\newcommand{\dfifty}{$\delta_{1950.0}$}
\newcommand{\lsun}{L$_{\odot}$}
\newcommand{\rf}{\par\noindent\hangindent 15pt {}}
\newcommand{\apj}[2]{ApJ #1, #2}
\newcommand{\aps}[2]{ApJS #1, #2} 
\newcommand{\am}[2]{A\&A #1, #2}
\newcommand{\as}[2]{A\&AS #1, #2} 
\newcommand{\ara}[2]{ARA\&A #1, #2} 
\newcommand{\aj}[2]{AJ #1, #2}
\newcommand{\mn}[2]{MNRAS #1, #2}
\begin{document}
\pagestyle{myheadings}
\markright{}

\begin{center}
{\Large\bf THE EXTENDED 12 MICRON GALAXY SAMPLE}

\vspace*{3cm}
Brian Rush and Matthew A. Malkan \\
Department of Astronomy, University of California at Los Angeles \\
Los Angeles, CA 90024--1562

\vspace*{.4cm}
Luigi Spinoglio \\
Istituto di Fisica dello Spazio Interplanetario, CNR \\
CP 27 00044 Frascati \\
Italy \\

\vspace*{3cm}
Accepted for publication in: {\it The Astrophysical Journal Supplement} \\
\end{center}

\begin{table}[h]
\centering
\begin{tabular}{rl}
Received: & 1992 January 13 \\
Revised:  & 1993 April 2 \\
Accepted: & 1993 April 13
\end{tabular}
\end{table}

\vfill\eject

{\centering\bf Abstract}

We have selected an all--sky ($|b|\geq25^\circ$) 12\um\ flux--limited sample of
893 galaxies from the {\it IRAS Faint Source Catalog, Version~2} (FSC--2). This
new sample contains 2.3 times as many objects as an earlier selection
(Spinoglio
\& Malkan 1989) based on the {\it IRAS Point Source Catalog, Version~2}. We
have
obtained accurate total fluxes in the IRAS wavebands by using the ADDSCAN
procedure for all objects with FSC--2 12\um\ fluxes greater than 0.15 Jy and
increasing flux densities from 12 to 60\um, and defined the sample by imposing
a
survey limit of 0.22 Jy on the total 12\um\ flux. Its completeness is verified,
by means of the classical $Log~N-Log~S$ and $V/V_{max}$ tests, down to 0.30~Jy,
below which we have measured the incompleteness down to the survey limit, using
the $Log~N-Log~S$ plot, for our statistical analysis. We have obtained
redshifts
(mostly from catalogs) for virtually all ($98.4\%$) the galaxies in the sample.

Using existing catalogs of active galaxies, we defined a subsample of 118
objects consisting of 53 Seyfert~1s and quasars, 63 Seyfert~2s, and 2
blazars--$\sim 13\%$ of the full sample), which is the largest unbiased sample
of Seyfert galaxies ever assembled. Since the 12\um\ flux has been shown to be
about one--fifth of the bolometric flux for Seyfert galaxies and quasars, the
subsample of Seyferts (including quasars and blazars) is complete not only to
0.30~Jy at 12\um\, but also with respect to a bolometric flux limit of $\sim
2.0\times 10^{-10}erg~s^{-1}cm^{-2}$. The average value of $V/V_{max}$ for the
full sample, corrected for incompleteness at low fluxes, is $0.51\pm0.04$,
expected for a complete sample of uniformly distributed galaxies, while the
value for the Seyfert galaxy subsample is $0.46\pm0.10$, suggesting that
several
more galaxies are yet to be identified as Seyferts in our sample. We have
derived 12\um\ and far--infrared luminosity functions for the AGN, as well as
for the entire sample. The AGN luminosity functions are more complete than
those
of the optically selected CfA Seyfert galaxies for all luminosities and AGN
types.

We extracted from our sample a complete subsample of 235 galaxies flux--limited
(8.3 Jy) at 60\um.  The 60\um\ luminosity function computed for this subsample
is in satisfactory agreement with the ones derived from the bright galaxy
sample
(BGS; Soifer et al. 1987) and the deep high--galactic latitude sample (Lawrence
et al. 1986), both selected at 60\um.  Over the high luminosity range where our
sample and the BGS overlap, however, our space densities are systematically
lower by a factor of $\sim 1.5$, whereas at low luminosities our space
densities
are higher by about the same amount. Comparable results are obtained when
comparing the far--IR luminosity function of our entire sample with the one
derived from the BGS. This is not unexpected, because of the bias towards
high-luminosity spirals caused by selection at 60\um.

{\it Subject headings:} galaxies: nuclei --- galaxies: Seyfert --- infrared:
sources --- luminosity function --- quasars --- galaxies: photometry

\subsection*{1. INTRODUCTION}

Complete and unbiased samples of active galactic nuclei (AGN) are essential
when
addressing the fundamental issues of the physical nature of galactic activity.
For example, as a function of bolometric luminosity, how do the space densities
of quasars, Seyfert~1~and~2 galaxies, radiogalaxies, LINERs and starburst
galaxies compare to those of normal spiral and elliptical galaxies? Are all
Seyfert~2s really dusty Seyfert~1s whose broad--line regions are completely
obscured? Is the apparently decreasing importance of dust in more luminous
active nuclei due to selection effects?  How many Seyfert~2s and dusty Seyfert
1s are missing from current catalogs? Complete samples are also needed for
statistical analysis leading to the major goal of minimizing the number of
parameters required for a physical explanation of all AGN.

Because the nonstellar processes powering AGN have many manifestations, these
objects span a diverse range of appearance, and it is nearly impossible to
obtain a complete sample of them for statistical studies. Thus, nearly all
previous AGN samples have suffered from some form of selection effects and/or
incompleteness, diminishing their usefulness in obtaining results
representative
of all AGN. For example, far--infrared surveys preferentially select dusty AGN,
whose bolometric luminosity is largely re--radiated by dust grains, whereas the
ultraviolet--excess searches of Markarian and Green (Green, Schmidt, \& Liebert
1987) are well known to be biased against reddened and dusty nuclei, in favor
of
blue Seyfert~1 nuclei and quasars.

We eliminate such difficulties by following the approach originated by
Spinoglio
\& Malkan (1989, hereinafter SM) -- a selection based on a flux limit at 12\um,
a waveband which minimizes wavelength--dependent selection effects. SM showed
that the 12\um\ flux carries an approximately constant fraction of the
bolometric flux (about one--fifth) for all types of Seyfert galaxies and
quasars. Now, by selecting at 12\um, we have obtained a sample of these objects
which is complete relative to a bolometric flux of $\sim 2.0\times
10^{-10}erg~s^{-1}cm^{-2}$. See SM for a more detailed discussion of this
approach.

In this paper we extend the 12\um\ galaxy sample selected by SM to a lower flux
limit (0.22~Jy compared to 0.30~Jy), using the {\it IRAS Faint Source Catalog,
Version 2} (Moshir et al. 1991, hereinafter FSC--2). To avoid the systematic
underestimates of the FSC--2 fluxes of all resolved sources, we obtained
accurate total flux measures for all the objects selected with the FSC--2,
using
the ADDSCAN procedure, with the help of the team at the Infrared Processing and
Analysis Center (IPAC). (See Appendix~B for a comparison of FSC--2 and ADDSCAN
fluxes) Two main objectives are achieved: first, completeness is verified down
to 0.3~Jy (as opposed to 0.5~Jy in SM); second, the number of objects is more
than double that in SM (893 compared to 390), allowing the production and
analysis of the largest unbiased sample of Seyfert galaxies (118, including
quasars and two BL Lac objects).

\subsection*{2. SAMPLE SELECTION}

We have selected all sources in the FSC--2 that meet the following criteria:
(1)
12\um\ flux density in the FSC--2 $\geq$ 0.15~Jy; (2) either $F_{60{\mu}m} \geq
1/2 F_{12{\mu}m}$ or $F_{100{\mu}m} \geq F_{12{\mu}m}$ (in order to exclude
most
stars and virtually no galaxies); (3) 12\um\ flux density from ADDSCANs $\geq$
0.22~Jy; and (4) $|b|\geq25^\circ$ (to avoid contamination from galactic
objects, especially stars).  In addition, the 12\um\ flux must have had a
moderate or high flux quality flag, as did the flux at 60 or 100\um, in order
to
assure real detections.

The details of our final selection of galaxies, and rejection of galactic
sources or galaxies contaminated by stellar objects, are given in Appendix~A.
The total number of sources in our present sample is 893, more than double
(229\%) our original 12\um\ sample. This includes 118 Seyfert galaxies and
quasars, exactly two times the 59 reported in SM.  The smaller increase in the
percentage of Seyferts is likely due to the fact that the fainter, more
distant,
sources in the new sample have not been studied in detail and thus probably
include undiscovered Seyferts.

\subsection*{3. COMPLETENESS}

In order to derive reliable statistical results, we had as our two highest
priorities obtaining a sample which was complete and yet large. We therefore
selected our flux limit to be 0.22~Jy, above the completeness limit of the
FSC--2, which is 0.18~Jy at 12\um.  However, it was necessary to obtain all
sources with a 12\um\ flux from the FSC--2 above 0.15~Jy in order to find those
objects with ADDSCAN whole--galaxy fluxes above our limit. The use of the total
fluxes, as compared to the FSC--2 fluxes, deteriorates the level of the
completeness limit of the sample from 0.2 Jy to 0.3 Jy. This happens for
several
reasons. First, while the number of sources remains constant, their fluxes will
systematically increase for any resolved source, so in a $Log~N-Log~S$
representation the points will be shifted towards higher fluxes. Second, the
FSC--2 is incomplete below 0.18~Jy, so we miss some sources which would have
cataloged fluxes below that level if the FSC--2 were complete to lower fluxes.
Third, there will be some sources with FSC--2 fluxes below 0.15~Jy, yet with
ADDSCAN fluxes still above 0.22~Jy. At lower FSC--2 flux levels, the
incompleteness grows while the number of objects with ADDSCAN fluxes above
0.22~Jy decreases. Thus we searched the FSC--2 only down to 0.15~Jy and then we
measured the resulting incompleteness of our sample.  A standard $Log~N-Log~S$
plot, shown in Figure~1, is fit well by a slope of $-3/2$ (the straight line),
as expected for a spatially uniform distribution of galaxies, down to 0.30~Jy,
while at lower flux levels the sample gradually becomes incomplete, reaching a
level of incompleteness of $\sim 40\%$ at 0.22~Jy. This shows marked
improvement
over the original 12\um\ sample of SM, which is complete only down to 0.50~Jy,
reflecting the greater incompleteness of the {IRAS Point Source Catalog,
Version~2} (1988, hereinafter PSC--2), from which that sample was selected.

By fitting a function which is exponentially decreasing towards lower flux
levels (the lower, curved line in Figure~1), we have estimated the
incompleteness of our sample between 0.22~Jy and 0.3~Jy. We then use this
estimate to correct the 12\um\ luminosity functions that we derive in
Section~8,
in the same manner as in SM.  We chose to correct for incompleteness only up to
0.30~Jy because of the good fit of the $-3/2$ slope line down to that flux, and
because any incompleteness above that level is not significant, being an
artifact of forcing an analytic smooth curve to the full range of data. In
Figure~2.{\it a}, we show the standard volume test of Schmidt (1968), corrected
for the incompleteness as estimated from Figure~1.  This gives an average
$V/V_{max}$ of $0.51 \pm 0.04$ for the entire sample.  For the computations of
the volumes ($V$) over which each galaxy is observed and the maximum volumes
($V_{max}$) over which it could be observed (given the flux limit of our
sample)
we have used redshifts which are corrected for solar motion within the Galaxy
and for a dipole Virgocentric flow, using the same correction as Geller \&
Huchra (1983), with an assumed infall velocity of $300~km~s^{-1}$.

\subsection*{4. FRACTIONS OF GALAXY TYPES}

A preliminary classification of galaxy nuclear activity has been made using
existing catalogs of active galaxies (SM; V\`eron--Cetty \& V\`eron 1991;
Hewitt
\& Burbidge 1991; the NASA/IPAC Extragalactic Database). About 13\% of the
galaxies in the sample are known to have Seyfert nuclei: 53 are Seyfert~1s and
quasars, 63 are Seyfert~2s, and 2 are blazars (OJ~287 and 3C~445). Including
the
29 LINERs brings the number of AGN in the sample to 16\%. Thirty--eight
non--Seyfert galaxies have an infrared luminosity in excess of
$6\cdot10^{44}~erg~s^{-1}$, probably due to violent star formation activity.
These might be considered active galaxies in a more general sense and we
classify them as {\it Starburst}\footnote{This classification is not identical
to other popular, optically defined, definitions of starburst galaxies; we use
it only for clarity when referring to this group of objects in this paper.}
galaxies. Including these objects would increase to 20.5\% the active galaxies
in our sample.

The Seyfert~1s, Seyfert~2s, Starburst galaxies and LINERs are listed in Tables
1--4, respectively, and the ``normal" galaxies (nearly all spirals) are given
in
Table 5.  These tables include, for each galaxy:  the name, the equatorial
coordinates (for the equinox 1950.0), the total IRAS flux densities measured by
the ADDSCAN procedure, the redshift, and the reference from which the redshift
has been taken.

Most ($>90\%$) of the flux densities quoted are those which are referred to as
Fnu(z) in the ADDSCAN data.  This represents the integral of flux between the
two zero--crossings of the continuum in the in--scan direction for the median
of
all scans.  We examined hundreds of addscans by eye to find this flux to be the
most accurate for normal cases.  For all other cases, specifically those
objects
for which the various ADDSCAN flux estimators differed significantly and/or for
which the FWHM were larger than 1.5 arcmin, we examined the ADDSCAN data to
determine the most accurate flux to use for each object at each waveband.  This
led us to sometimes use Fnu(t) (the flux between two points at a fixed distance
from the center coordinate) or the template flux (fit to the median scan of
each
object) instead of Fnu(z). Finally, for the most extended objects in our sample
we adopted the fluxes from Rice et al.'s Catalog of IRAS Observations of Large
Optical Galaxies (1988; hereafter LGC). For nearly all of the objects, unless
explicitly specified in the tables footnotes, the redshift given is the
observed
one, before any corrections.

To determine the completeness of our sample for different classes of galaxies,
the standard volume test is presented separately for each class in Figure~2{\it
b}, corrected as in Figure~2{\it a}. We did not consider our starburst galaxy
and LINER samples in this graph because they are not complete subsamples. The
``starbursts" are incomplete because they are a luminosity--limited sample by
definition. Similarly, the LINERs do not form a complete subsample because
their
identification among our full sample has not been done systematically by
examining the optical spectra, but only by a literature search, and therefore
it
is highly biased towards the northern hemisphere and towards the brightest and
nearest (low z) objects. The values of $V/V_{max}$ for the Seyferts in the
sample are below that of the non--Seyferts, most likely because a number of
faint and distant Seyfert~1s and~2s in our sample have not yet been identified
as such.

\subsection*{5. SKY AND REDSHIFT DISTRIBUTIONS}

An all--sky plot of the entire sample (Figure~3) shows the galaxies are
distributed roughly randomly over the $|b|\geq25^\circ$ sky.  The most
noticeable departure from this is the concentration of Virgo--cluster galaxies
around \afifty = $12^h20^m$, \dfifty = $+20^\circ$, centered on M87. Although
the Seyferts alone show only a slight imbalance of more objects to the south
(see Figure~3), 26 of the 29 LINERs in our sample are at positive declinations.
This most likely results from the more intensive spectroscopic investigation of
northern galaxies, compared to those in the south. In addition, the fact that
most of the LINERs in our sample are nearby (average corrected redshift is
0.003) indicates that a number of more distant LINERs have not yet been
identified as such. If so, we can expect the actual number of LINERs to be much
greater, bringing our sample in closer agreement with previous studies which
have suggested a higher ratio of LINERs to Seyferts (Heckman 1980; Woltjer
1990).

We corrected the observed redshifts for Galaxy rotation and motion towards the
Virgo cluster using the same correction adopted in Geller \& Huchra (1983). We
had to ensure that the redshifts we used represent cosmological motion, and are
thus proportional to distance. We therefore took the mean flow--corrected
galactocentric group velocity for each of the 109 galaxies in our sample which
are members of a CfA group (Geller \& Huchra 1983), and for which the
difference
between individual and group velocity is less than $500~km~s^{-1}$. This
approach eliminates the contributions of peculiar motion within the groups,
while ensuring that only minor errors in the luminosities could result from
possible incorrect group associations. Throughout this work, we assumed $H_o =
75~km~s^{-1}Mpc^{-1}$.

The histograms showing the distribution of distances among differ\-ent
class\-es
of galax\-ies are presented in Fig.~4, where the normal galaxies are compared
to
Seyfert~1's and LINERs (top), and to Seyfert~2's and starburst galaxies
(bottom). The average corrected redshift of the sample is 0.013 ($cz =
3858~km~s^{-1}$), giving an average distance of 51 Mpc. The Seyfert~1 galaxies
and quasars are the more distant objects in the sample, with an average
redshift
of 0.035, followed by the starburst galaxies at 0.030.  This is not surprising
for the latter, which were defined as high luminosity galaxies. The Seyfert~2s
have an average redshift of 0.021, and the LINERs 0.0029.  Thus the average
redshift of all {\it active} galaxies is 0.028 ($8419~km~s^{-1}$), much greater
than the 0.013 ($3858~km~s^{-1}$) found for normal galaxies (that is all
galaxies which are not Seyfert~1s or~2s, quasars, starburst galaxies, or
LINERs)
in the sample.

\subsection*{6. INFRARED COLORS AND SLOPES}

The color--color diagram showing the distribution of $Log
(F_{60{\mu}m}/F_{25{\mu}m})$ vs. $Log (F_{25{\mu}m}/F_{12{\mu}m})$ is shown in
Fig.~5{\it a} for Seyfert galaxies and in Fig.~5{\it b} for starbursts and
LINERs, both compared to normal galaxies. There is a clear tendency for the
Seyfert galaxies to have flatter slopes from 25 to 60\um.  Although this is the
best infrared color discriminator between Seyfert and normal galaxies, it is
not
decisive, since about 40\% of the Seyferts are found among the normal galaxies
in this plot. More precisely, the color selection criterion used to select the
``warm extragalactic objects'' by Low et al. (1988) ($F_{25{\mu}m}/F_{60{\mu}m}
> 0.25$ -- see the dashed line in Fig.~5{\it a}), if applied to our sample,
would have missed 20 Seyfert~1s (38\%) and 29 Seyfert~2s (46\%). Their
criterion
would also have excluded all but one of the high $L_{FIR}$ galaxies in our
sample, as shown in Fig.~5{\it b}.  A discriminator based on a high
25--to--12\um\ flux ratio would better select these latter objects.  The
vertical line in Fig.~5{\it b} (corresponding to $F_{25{\mu}m}/F_{12{\mu}m} >
4.79$), for example, separates 14 of the 38 high $L_{FIR}$ galaxies from 97\%
of
the normal galaxies.

The LINERs, on the other hand, have average colors very similar to those of the
normal early--type galaxies in our sample, but with an even lower average value
of $Log~(F_{25{\mu}m}/F_{12{\mu}m})$. This indicates that much of the 12\um\
emission does {\it not} come from star--forming regions (which produce more
flux
at 25\um\ than at 12\um).  These colors therefore imply that our 12\um\
selection is not as successful in detecting LINER active nuclei as it is for
Seyfert nuclei.

To investigate the extent to which beam resolution may have affected the
observed infrared colors of the sample galaxies, we have examined the
difference
between several spectral slopes derived from the FSC--2 fluxes and the same
slopes as derived from the ADDSCAN fluxes. While we refer to Appendix~B for the
details of the analysis and the plots, we note here that no explicit effects of
beam size on infrared colors is apparent for the (more point--like) Seyfert
galaxies in our sample, and only small effects are seen among the normal
galaxies.

\subsection*{7. COMPARISON WITH OTHER GALAXY CATALOGS}

\subsubsection*{7.1. {\it The CfA--Seyfert and Bright Galaxy Samples}}

Figure~6 is a Venn diagram showing the overlaps among our sample, the IRAS
Bright Galaxy Sample (hereinafter BGS, Soifer et al. 1987, 1989), and the CfA
Seyferts ({\it e.g.}, Edelson, Malkan, \& Reike 1987). Our sample has 268
(86\%)
of the 313 objects in the BGS.  The other 45 BGS objects are among the fainter
galaxies in that sample, all having either a 12\um\ flux below our completeness
limit of 0.30~Jy or upper limits only in the FSC--2. To compare our sample and
the BGS further, we defined subsamples from each. Our complete 60\um--selected
subsample (at 8.31~Jy; see \S~8 for further details) has 235 galaxies.  Of the
174 which are in the BGS' area of sky, 172 are in the BGS. Conversely, of the
122 galaxies in a complete 12\um--selected subsample of the BGS (at 0.79~Jy;
Soifer \& Neugebauer 1991), 116 are in our 12\um Sample (the other 6 having
upper limits in the FSC--2).  This similarity indicates that the two methods
would select virtually identical lists of galaxies down to 0.79~Jy at 12\um\ or
8.31~Jy at 60\um\.  At lower flux levels however, 12\um-- or 60\um--selection
will include relatively more galaxies with flat or steep IRAS spectra,
respectively. See \S~8.4 on how this effect, combined with the correlation of
IRAS color with luminosity, will affect luminosity functions derived from the
two samples.

To compare specifically the Seyfert galaxies in these samples, we note that our
12\um\ Sample includes 13 of the 27 Seyfert~1s and 8 of the 21 Seyfert~2s from
the CfA sample, whereas the BGS contains only 5 and 7 of these 12\um/CfA
Seyfert~1s~and~2s, respectively, and no other CfA Seyferts, although all the
CfA
Seyferts are within the sky area surveyed by both the 12\um\ and the BGS
samples. The BGS also contains 50\% (20/40) of the 12\um\ sample Seyfert~2s
which are within their surveyed sky area, but only 26\% (9/34) of our
Seyfert~1s
in that area. (The fraction of non-Seyferts is slightly less (42\%, or 239/546)
than the fraction of Seyfert~2s.) Although the BGS covered less sky area
($|b|\geq 30^\circ$, plus restrictions on declination) than our sample, this
cannot account for all of the missing Seyferts. More importantly, the BGS'
deficiency of Seyfert~1s {\it relative} to Seyfert~2s as compared to the 12\um\
sample supports SM's claim that far--infrared (e.g. 60\um) surveys are biased
against AGN with hotter colors, such as Seyfert~1's.

\subsubsection*{7.2. {\it The 6~cm Northern Sky Catalog}}

We have compared our entire list of galaxies with the 6~cm Northern Sky Catalog
(Becker, White, \& Edwards 1991, hereinafter BWE). That survey detected 30 of
the 52 12\um\ Seyferts (58\%) in the area of sky covered but only 113 (27\%) of
the 419 non--Seyferts. This is not surprising, as Seyfert galaxies often have
strong radio emission. Figure~7 is a plot of 6~cm luminosity (from BWE) against
60\um\ luminosity for the 471 northern galaxies. Upper limits given for those
objects not in BWE are the (declination--dependent) flux--limits reported in
that work.

To quantify the correlation between $\log L_{6cm}$ and $\log L_{60\mu m}$ in
our
sample, we have used the ASURV software package (Isobe, La~Valley, \& Feigelson
1992). ASURV provides two statistical tests for the presence of a linear
correlation between two variables when one of these variables ($\log L_{6cm}$)
is heavily censored (mostly upper limits). Both the Cox Proportional Hazard and
Generalized Kendal's Tau tests indicate that the correlation between $\log
L_{6cm}$ and $\log L_{60\mu m}$ is significant at the 99\% level. We used
Schmitt's Linear Regression and the Buckley-James method to estimate the slope
of this correlation. Both methods give similar results:
$\log L_{6cm}\sim Constant\times\log L_{60\mu m}^{1.0}$ for the non-Seyferts
and
$\log L_{6cm}\sim Constant\times\log L_{60\mu m}^{1.2}$ for the Seyferts,
with similar values for the constants. Since the two regression techniques
agree
well with each other and indicate that the relation is virtually linear, we
then
assumed a direct proportionality.  This problem then becomes univariate, where
the single variable is the ratio of radio to IR flux, which is for many
galaxies
only an upper limit. (Clearly there is real scatter in the relation, but this
will average out, as long as the linear relation holds). In the present case we
cannot assume we know the functional form of this distribution of radio/IR flux
ratios. Therefore we use the Kaplan-Meier product-limit estimator (also part of
the ASURV software), which is known to be a maximum-likelihood indicator, even
when most of the measurements are censored (upper limits; Feigelson \& Nelson
1985). The KM estimator of the luminosity ratio distribution is defined only at
the values of the actual detections.  The upper limits are then re-distributed
uniformly among all the bins of lower detected values. Of course upper limits
which are higher than the highest detection have no weight in this process. The
KM estimator yields a median of $\log(L_{6cm}/L_{60\mu m}) = -4.98\pm 0.13$
(upper dotted line) for the Seyferts and $-5.61\pm 0.05$ (lower dotted line)
for
the non--Seyferts.

The solid line shown for comparison represents a constant value of \linebreak
$\log (L_{6cm}/L_{60\mu m})$ $=-5.64$. This number is obtained by using the
average value of $Q_{60}$ ($\equiv f_{\nu=60\mu m}/f_{\nu=20cm}$) given in
Bicay
\& Helou (1990) for a sample of 25 normal spiral galaxies, and assuming a
$-0.7$
slope power law from 6~to~20~cm. This line in identical (within error) to that
calculated for our non--Seyferts.

Much of the scatter in this plot may be attributable to low--level nuclear
activity. The radio--detected non--Seyferts are nearby ($<cz> =
2520~km~s^{-1}$), as compared to the undetected non--Seyferts ($<cz> =
3390~km~s^{-1}$). Many of these radio emitting non--Seyferts are likely to
harbor some form of nuclear activity. For example, 54\% of our LINERs in the
area surveyed by BWE were detected by them. The higher average ratio of 6~cm to
60\um\ luminosity for Seyferts is consistent with previous claims that Seyferts
have excess radio power as compared to normal galaxies in the infrared--radio
correlation (Condon 1992). This ratio might even be used to distinguish AGN
from
starbursts since the correlation has been found to be much tighter for the
latter (Condon et al. 1982).

\subsection*{8. LUMINOSITY FUNCTIONS}

\subsubsection*{8.1. {\it Derivation}}

The 12\um\ and the far--infrared luminosity functions (hereinafter LFs) have
been calculated for the entire sample, as well as for subsamples of the
different galaxy types. We have also extracted a complete 60\um\ subsample, in
order to correctly derive the 60\um\ LF and compare it to previous works. For
the Seyfert galaxies alone, we also computed the 60\um\ LF from our entire
sample and compared it to the one derived for the optically selected CfA
Seyferts. Although neither of these latter subsamples are complete at 60\um,
they represent the best available optically and infrared selected samples of
Seyfert galaxies, making a comparison meaningful.

All LFs have been derived using Schmidt's (1968) estimator
$$\Phi={4\pi\over \Omega} \sum{1\over V_{max}},$$
where $V_{max}$ was individually computed for each galaxy in the sample.  The
luminosities incorporate the small K--correction (Sandage 1975) calculated
assuming each IRAS point can be connected by a piecewise power--law. The
redshifts have been corrected for solar motion as described in Section~5.

We note here that we have corrected the two errors in SM that affected
calculations of the LFs. A typographical error in the equation correcting
observed redshifts for solar motion led to incorrect values of volumes and
luminosities by an average of a few percent. More substantially, in SM all the
LFs were underestimated by a constant factor of 2.5, which affected some of the
conclusions drawn from them.

\subsubsection*{8.2. {\it 12\um\ Luminosity Functions}}

The 12\um\ space densities for Seyfert~1~and~2 galaxies and for the entire
sample are given in Table 6. Figs.~8{\it a}~and~{\it b} show the 12\um\ LF for
Seyfert galaxies compared to non--Seyferts and for Seyfert~1s compared to
Seyfert~2s, respectively. Again the quasars and blazars have been grouped
together with Seyfert~1s. We fit our LFs to a function involving two power
laws,
after that used in Lawrence et al. (1986).  The analytical form fitted is
$$\Phi(L)=CL^{1-\alpha} \Bigl( 1+{L\over L_{\star}\beta} \Bigr) ^{-\beta}.$$
As expected, normal galaxies are the most numerous objects at the lower
luminosities, while Seyfert galaxies begin to dominate the space density around
$10^{10.8}$ \lsun, and all galaxies found above $L_{12{\mu}m}\sim10^{11.2}$
\lsun harbor active nuclei. The sharp turnover seen in the non--Seyfert LF
(indicated also by the high value of $\beta$, the change in slope from low to
high luminosities) resembles the ``knee" observed in optical luminosity
functions of normal galaxies. The best--fit parameters are:
$$\alpha=1.3,~\beta=2.1,~\log L_{\star}=9.8~~\mbox{(Seyferts)}$$
and
$$\alpha=1.7,~\beta=3.6,~\log L_{\star}=9.8~~\mbox{(non-Seyferts).}$$
In Fig.~8{\it b} we added the constraint $\alpha\equiv1.0$. This affects only
the fits at the lowest luminosities, where small statistics make the results
uncertain. Otherwise, the individual fits show the luminosity function of Sy~2s
to be slightly higher than that of Seyfert~1s, except that Seyfert~1s extend to
higher luminosities. The best--fit parameters are:
$$\beta=2.1,~\log L_{\star}=9.2~~\mbox{(Seyfert 1s)}$$
and
$$\beta=2.5,~\log L_{\star}=9.6~~\mbox{(Seyfert 2s).}$$
In Fig.~8{\it c}, we also show the space densities we derived from our samples
of LINERs and starburst galaxies as compared to Seyfert and normal galaxies.
For
the LINERs these densities must be considered as lower limits, because of the
known incompleteness of the sample (see \S~V). Both the fraction of LINERs
among
the entire sample and the ratio of the LINER to the normal galaxy populations
strongly decrease with luminosity. This decrease is expected if the number of
LINERs in our sample not yet identified increases with redshift, and if the
intrinsic luminosity gain is less than the loss due to the distance. Because
our
starburst galaxy sample has been defined merely by imposing a lower far--IR
luminosity limit to the non--Seyfert population, as mentioned above, the
corresponding points in Fig.~8{\it c} cannot be considered to represent the
starburst LF at low luminosities. The cut--off seen below
$L_{12{\mu}m}=10^{43.9}~erg~s^{-1}$ is of course due to our selection
definition. At higher luminosities, however, the sample becomes complete, and
it
appears that the starburst (i.e. non-Seyfert) space density, while dominating
at
$L_{12{\mu}m}=10^{44.1}~erg~s^{-1}$, falls below the Seyfert space density at
$L_{12{\mu}m}=10^{44.5}~erg~s^{-1}$ by more than one order of magnitude.

A comparison between our 12\um\ luminosity function for the whole sample and
the
one derived by Soifer \& Neugebauer (1991) by extracting a complete 12\um\
subsample of 122 galaxies from the BGS results in a very good agreement in the
luminosity range $10^8$\lsun$<LogL<10^{11}$\lsun, where their sample has good
statistics.

\subsubsection*{8.3. {\it Far--infrared Luminosity Functions}}

The far--infrared luminosities have been computed by integrating the spectral
energy distributions over the four {\it IRAS} wavebands, covering the
wavelength
range from 12 to 100\um. Although it is not strictly correct to define a LF in
the whole far--IR, where our sample may not be complete, this can give a
first--order estimate of the bolometric LF, especially for normal galaxies. The
LF for the full sample is similar to that derived by Soifer et al. (1987) for
the 60\um--selected BGS.  (In the following section we compare the better
defined 60\um\ LFs of the two samples). The space densities for each class of
Seyfert and for the entire sample are listed in Table 7. Figs.~9{\it
a}~and~{\it
b} compare the far--IR LFs of Seyfert galaxies in our sample with
non--Seyferts,
and with every galaxy class, respectively. As in the case of the 12\um\ LF, the
space densities given for LINERs are strictly lower limits, and the space
density at the lower luminosity bin for ``starburst" galaxies is truncated by
the selection definition. The behavior at higher far--IR luminosities is
similar
to that of the 12\um\ LF, confirming the fall-off of the starburst galaxies as
compared to the Seyferts by $L_{FIR}\simeq 10^{45.3}~erg~s^{-1}$. Moreover, the
most luminous far--IR galaxies in our sample are still AGN, there being 4
Seyfert~1s/quasars, 6 Seyfert~2s, and 1~BL~Lac more luminous than any
non--active galaxy in this range. This difference would appear even greater
with
a truly bolometric LF, which would give more weight to the Seyferts over the
starburst galaxies, since the former are brighter at virtually all
non--infrared
wavelengths. The difference of our conclusion from that of Soifer et al. (that
infrared ultraluminous galaxies are the ``dominant" population of emitters in
the universe) reflects the fact that far--IR selected surveys, including the
one
they did at 60\um\, miss most of the bluer (i.e. flatter IR slope) Seyfert~1
galaxies and quasars which our 12\um\ selection efficiently detects.
Nevertheless, a definitive statement can not yet be made about the shape of the
starburst LF at high luminosities, since there are too few galaxies to specify
how it continues from the LF at lower luminosities.

\subsubsection*{8.4. {\it 60\um\ Luminosity Functions}}

To compare rigorously the 60\um\ LF from our sample with those of the
60\um--selected BGS (Soifer et al. 1987) and of the high--galactic latitude
60\um\ sample (Lawrence et al. 1986), we have extracted from our 12\um\ sample
a
subsample of galaxies complete at 60\um. To find the limiting flux density at
60\um\ for which our 12\um\ sample is complete, we plotted in Fig.~10 the
differential number counts as a function of the 60\um\ flux. As can be seen,
the
expected slope of $-3/2$ fits the number counts down to a 60\um\ flux density
of
$\sim 8.31~Jy$. This is in good agreement with what one would expect, since the
reddest objects in our sample have $F_{60{\mu}m} \approx 30 \cdot
F_{12{\mu}m}$,
and $30 \cdot 0.3~Jy$ (our 12\um\ completeness limit) then predicts a 60\um\
completeness limit of about 9~Jy. Thus we defined a complete flux--limited
sample of galaxies that contains 235 objects. The 60\um\ LF of this subsample
is
given in Table 8. In Fig.~11 we have plotted our 60\um\ LF along with those of
the BGS and of sample~3 of Lawrence et al. The latter is the weighted
combination of their `main sample' ($b>60^{o}$, $0<l<110^{o}$,
$S_{60{\mu}m}\geq
0.85 Jy$) and their `smaller' sample ($b>60^{o}$, $0<l<110^{o}$,
$RA<13^{h}~45^{m}$, $0.5~\leq S_{60{\mu}m} < 0.85 Jy$), converted to the units
and value of $H_o$ adopted here. We fitted each sample to the function given in
Lawrence et al. The best--fit parameters are:
$$\alpha=1.7,~\beta=1.8,~\log L_{\star}=10.1~~\mbox{(60\um\ subsample),}$$
$$\alpha=1.5,~\beta=1.5,~\log L_{\star}=10.1~~\mbox{(Lawrence et al.),}$$
and
$$\alpha=1.1,~\beta=2.1,~\log L_{\star}=9.5~~\mbox{(Soifer et al.).}$$
Our 60\um\ space densities at mid and high luminosities are systematically
lower, presumably because of slight incompleteness in our sample
below\linebreak
$f_{\nu=60\mu m}\sim 9~Jy$ (as indicated by the value of $V/V_{max}\sim 0.45$
for our 60\um\ subsample at high luminosities) and our different selection
wavelength. A comparison with the LF derived by Lawrence et al. shows that,
while at low luminosities ($L_{60{\mu}m}/ L_{\odot} \leq 10^{10.5}$) the two
LFs
are consistent, at high luminosities the deep 60\um\ sample shows even higher
space densities than the BGS.  At $L_{60{\mu}m}~=~10^{8}$\lsun\, the ratios of
our fitted LF to those of Soifer et al. and Lawrence et al. are 1.59:1.23:1.0,
respectively, whereas, at $L_{60{\mu}m}=10^{11}$\lsun\, the same ratios are
1.0:1.16:1.75.  These trends in the luminosity functions are not surprising
since more luminous spiral galaxies emit a systematically larger fraction of
their total luminosity at 25 and 60\um\ (e.g., Soifer \& Neugebauer 1991;
Spinoglio et al. 1993). Therefore, sample selection at 60\um\ is relatively
better tuned than selection at 12\um\, for detecting galaxies with unusually
large far-infrared luminosities. To show this effect explicitly, we plot the
60--12\um\ color as a function of 60\um\ luminosity in Figure~12. This plot
shows that, excluding Seyferts and LINERs, cooler IRAS colors are correlated
with higher luminosities (a plot of the same color vs. Far-IR luminosity looks
almost identical). Therefore a sample with more hot/cool objects will tend to
have more low/high--luminosity galaxies, hence the difference between the two
samples. We note, however, that the 3~highest luminosity bins in Figure~11,
where these differences are seen, represent just over $\sim$10\% of the
samples,
and the lowest luminosity bins include an even smaller fraction.

We have also compared the 60\um\ LF of the Seyfert galaxies in our entire
sample
to that of the optically selected CfA sample (see, for example, Edelson,
Malkan,
\& Rieke 1987).  These 48 Seyferts, spectroscopically selected from the CfA
redshift survey (Huchra et al. 1992), were thought to comprise a complete and
unbiased AGN sample. But the survey was based on properties of the host galaxy,
and thus was not genuinely flux--limited with respect to the active nucleus. In
Fig.~13{\it a}~and~{\it b} we compare the 60\um\ LFs of the two samples for all
Seyfert galaxies together and for each Seyfert type, respectively. Fig.~13{\it
a} indicates higher space densities of Seyferts both at low ($L_{60{\mu}m} \leq
10^{44}~erg~s^{-1}$) and high ($L_{60{\mu}m}\geq 10^{45}~erg~s^{-1}$)
luminosities in our sample. In other words, our luminosity function more
closely
resembles a single power--law, whereas that of the CfA Seyferts turns over more
strongly in the mid--luminosity range. While the greater number of
high--luminosity objects is due to the inclusion among our Seyferts of the few
quasars and blazars in the 12\um\ sample, the higher density at lower
luminosity
indicates that our selection produces a more complete sample than the one
extracted from the CfA redshift survey. The same is true for both the
Seyfert~1s
and Seyfert~2s separately, as can be seen in Fig.~13{\it b}.  This figure also
shows that we found a result similar to the one of the CfA sample:  over most
of
the luminosity interval, the Seyfert~2s are more numerous than Seyfert~1s by
about a factor of 1.5 -- 2.

\subsection*{9. SUMMARY AND CONCLUSIONS}

The main results of this work can be summarized as follows:

1. We have defined a large sample of galaxies (893) using the 12\um\ flux as
given from the {\it IRAS} FSC--2. This new sample contains more than twice as
many objects as our earlier sample (SM), and its completeness is verified down
to 0.3~Jy. Because the 12\um\ flux is representative of the bolometric flux for
active galaxies (SM), we have obtained an AGN sample complete with respect to a
bolometric flux limit of $\sim 2.0\times 10^{-10}erg~s^{-1}cm^{-2}$.

2. About 13\% of the galaxies in the sample are known to have Seyfert or quasar
nuclei.  Another $\sim~3\%$ of the sample galaxies are classified as LINERs,
and
another $\sim~4\%$ have very high ($> 6 \cdot 10^{44}~erg~s^{-1}$)
far--infrared
luminosities, but not a Seyfert nucleus. Therefore about one fifth of the
sample
are galaxies that are ``active'' in a broad sense.

3. The high luminosity tail ($L_{FIR}>10^{45.3}erg~s^{-1}$;
$L_{12{\mu}m}>10^{44.5}erg~s^{-1}$) of the far--IR and 12\um\ luminosity
functions is dominated by Seyfert galaxies and quasars, and not by the high
$L_{FIR}$ nuclei.

4. The 60\um\ luminosity function, that we derive by extracting from our entire
sample a subsample complete at 60\um, is similar those derived from 60\um\
selected samples (Lawrence et al. 1986; Soifer et al. 1987), subject to the
slight tendency for 60\um\ to select non-Seyfert galaxies of high far-infrared
luminosity, relative to 12\um--selection.

5. A comparison of the 60\um\ luminosity function for all our Seyfert galaxies
with the one derived for the CfA Seyfert galaxies shows our Seyfert sample to
be
more complete, by a factor of $\sim 50\%$.  We find a slightly higher space
density of Seyfert~2s than Seyfert~1s, in agreement with the CfA results.

We are currently combining multiwavelength observations of the 12\um\ Galaxies
to construct {\it bolometric} luminosity functions, which will be published in
a
forthcoming article.

We thank Michael Strauss and Will Saunders for generously supplying redshifts;
Deborah Levine, Tim Conrow, Tom Soifer, George Helou and the staff at IPAC
(Caltech, Pasadena) for help in obtaining and understanding the IRAS data; and
Lick Observatory for allocating the time used to obtain the needed spectra.
This
research has made use of the NASA/IPAC Extragalactic Database (NED) which is
operated by the Jet Propulsion Laboratory, California Institute of Technology,
under contract with the National Aeronautics and Space Administration.  This
work was supported by NSF grant AST 85--52643 and NASA grants NAG--1358,
NAG--1449, and NAG--1719.

\subsection*{APPENDIX A}

\subsection*{DETAILS ON SOURCE INCLUSION AND REJECTION}

We originally based our sample on the FSC--2 12\um\ flux density. Applying our
selection criteria to the FSC--2 produced a list of 1105 objects with a 12\um\
flux density $\geq .20~Jy$. Following SM, we then excluded all objects in the
region of the Large Magellanic Cloud (109). The sources meeting our criteria
within the Small Magellanic Cloud (14; Schwering \& Israel, 1989) and M33 (5)
regions, which are associated with dark clouds, $H\alpha$ emission nebulae,
stellar cluster or HII regions  have also been excluded. In addition, 160 stars
found in the SAO (1966) and BSC (Hoffleit 1982) star catalogs were excluded, as
well as seven planetary nebulae, six of which were clearly identified as such
on
the POSS plates, and one identified by Strauss (1991).

We searched major extragalactic catalogs (e.g., V\`eron--Cetty \& V\`eron 1991;
Huchra et al. 1992 (CfA Redshift Survey; ZCAT); de~Vaucouleurs, de~Vaucouleurs,
\& Corwin 1976 (RC2); Lauberts 1982 (ESO/Uppsala Survey)) to find redshifts and
identifications. These supplied over 700 redshifts, and other sources (e.g.,
Saunders 1991; Strauss et al. 1990, 1992; Strauss 1991) provided about 60 more.
For all the remaining objects without known redshifts, we made 5x magnified
photographs from the POSS and ESO plates. From these finding charts, an
additional 45 objects, listed in Table 9, were shown to be clearly galactic
(mostly stars). This was further supported by their {\it IRAS} colors, which
did
not fall near the range established by the sample galaxies. We then obtained
optical spectra from the Lick Observatory 1m and 3m telescopes for virtually
all
the remaining objects.  These spectra provided us with several more redshifts,
and enabled us to exclude 19 more objects (listed in Table~10) as stars or
galactic nebulae. An additional 10 objects were excluded from our sample
because, although there is a galaxy close to the {\it IRAS} coordinates,
ADDSCANs obtained from IPAC, along with finding charts, indicated that the {\it
IRAS} object is actually a nearby star and/or that the galaxy's contribution to
the 12\um\ flux is less than 0.2 Jy, the rest coming from the star. One more
object (NGC~3395/6) is a double system, resolved by the {\it IRAS} ADDSCANs,
and
each galaxy contributes less than 0.2 Jy to the total 12\um\ flux. These 11
objects are listed in Table 11. Finally, we had to exclude (for now) three
other
objects because no redshift information was available. These are given in Table
12, together with their {\it IRAS} FSC--2 fluxes and comments. Thus,
altogether,
we excluded 373 objects from our original list.

To be as complete as possible, we also applied our selection criteria to the
entire {\it IRAS Faint Source Database} (FSDB), which contains all band--merged
sources extracted from the {\it IRAS Faint Source Survey} data (Moshir et al.
1991). This includes all the objects in the FSC--2, as well as those placed
into
the {\it Faint Source Reject File} (FSR). Although great attention went into
compiling the FSC--2, it is known that a few true sources ended up in the FSR.
We wanted to make sure that our sample did not exclude the galaxies in this
group. We thus applied our selection criterion to the FSDB. This search yielded
only 9 more sources, further indicating the completeness of the FSC--2. And six
of these, including the Seyfert~1 galaxy NGC~4151, were excluded from the
FSC--2
because they were in an area of sky not covered often enough by the {\it IRAS}
satellite to give the minimum necessary hour--confirmations for inclusion ({\it
IRAS Explanatory Supplement 1988}; Moshir et al. 1991).  We included the LMC,
the SMC and M~33 in our sample, using the {\it IRAS} fluxes given in the LGC,
since these galaxies are associated with many point--like sources in the
FSC--2.
In addition, we added the three galaxies (NGC~2992, NGC~5595 and NGC~5597) from
the SM sample for which the FSC--2 gives upper limits at both 60 and 100\um,
but
good detections at 12 and 25\um, while the PSC--2 fluxes are all good
detections. The number of galaxies included from sources other than the FSC--2
is thus 15, giving a total of 747 objects in the FSC--2--based extended 12\um\
sample.

Objects which are extended enough to be resolved by the IRAS beam will have
their FSC--2 flux densities underestimated, and thus ADDSCAN data is required
to
obtain accurate whole--galaxy fluxes. (See Appendix~B for graphs which show
this
explicitly for our entire sample.)  We have thus obtained complete ADDSCANed
flux information from IPAC for all objects with FSC--2 12\um\ flux densities
$\geq .15~Jy$.  We increased the limit of our sample to 0.22~Jy and added to
our
list 207 objects with an ADDSCAN 12\um\ flux density above this limit which
were
not in our list of 747 galaxies selected by criteria based solely on FSC--2.
Ten
other such objects might have been added, but have instead been placed in
Table~12 for now because we could find no redshift information for them. We
also
excluded 61 objects which were in the FSC--2 list of 747, but which have
ADDSCAN
12\um\ flux densities {\it below} 0.22~Jy (a few objects will actually have
lower ADDSCAN than FSC--2 fluxes because of limitations in the accuracy of the
FSC--2).  Thus, the FINAL count of objects in this sample is (747+207--61=) 893
galaxies, plus the 13 objects in Table~12 which are likely to be galaxies, but
for which we have not yet obtained a redshift.

In all of the calculations in this paper we use the ADDSCAN fluxes in each
waveband for most objects.  For the others, the LGC includes the largest
optical
galaxies that would have been resolved by the IRAS beam, causing their fluxes
to
be the most seriously underestimated. For 36 galaxies in our sample which are
also in the LGC (in addition to the LMC, SMC, and M33) we use the larger flux
densities from the LGC, which more accurately represent the flux from the
entire
galaxy, when making calculations.

\subsection*{APPENDIX B}

\subsection*{COMPARISON OF ADDSCAN AND FSC--2 FLUXES AND COLORS}

\subsection*{1. ADDSCAN--TO--FSC--2 FLUX RATIOS VS. REDSHIFT}

Since both the PSC--2 and the FSC--2 are point--source targeted surveys, they
underestimate the flux densities of resolved sources ({\it IRAS Explanatory
Supplement 1988}, Moshir et al. 1991).  To study this effect quantitatively, we
graphed the ratio of ADDSCAN--to--FSC--2 flux densities as a function of
redshift in each of the four {\it IRAS} wavebands.  Figures~B1{\it a}--B1{\it
d}
show the results for the normal galaxies in our sample and Figures~B2{\it
a}--B2{\it d} for the Seyfert Galaxies.

We binned the objects according to redshift (with bin sizes of $500~km~s^{-1}$
for the normal galaxies and $1000~km~s^{-1}$ for the Seyferts), and then
plotted
the median and first and third quartile points. The points in higher--redshift
bins are decreasingly reliable statistically, each bin above
$\sim7500~km~s^{-1}$ representing only a few objects. At 12\um, it can be seen
that the ADDSCAN flux densities are greater by a factor of about 1.6 for most
objects, with this ratio decreasing with redshift to a minimum value of about
1.2 for Seyferts and 1.3 for normal Galaxies. At 25\um, this same effect
exists,
but to a lesser degree, the ratio becoming as low as 1.1 at higher redshifts.
At
60 and 100\um\ the ratio is only $\sim 1.2$ for the most resolved sources and
decreases quickly to 1.05---1.1 for normal Galaxies and to 1---1.05 for
Seyferts. In each waveband, the effects are less for Seyferts than for normal
galaxies because they are more point--like.  These effects would be even less
for very point--like sources such as higher--redshift active galaxies.

One might expect these flux ratios on average to approach unity at sufficiently
high redshift, but there are a few reasons why this may not quite happen in
practice. First, since ours is a relatively low--redshift sample, even some of
the more distant sources could still be resolved by the IRAS beam, thus having
their flux densities under--estimated in the FSC--2.  Secondly, since our top
priority is to obtain estimates of the {\it total} galactic flux, we usually
use
the flux density referred to as Fnu(z) in the ADDSCANs. This is the integral of
flux between the two zero--crossings of the continuum in the in--scan direction
for the median of all scans.  At the lowest signal/noise ratios, this value
will
be slightly biased towards higher fluxes since, as the integral is calculated
outward from the peak position, negative noise in the source wings tends to be
neglected if it causes the signal to fall below the continuum (thus determining
the zero--crossing), whereas all positive noise fluctuations before the first
zero--crossing are included.  In a plot of (ADDSCAN/FSC--2) 12\um\ flux ratio
vs. SNR, this ratio tends to increase as SNR decreases below $\sim 10$. As a
test, we fitted the average of this trend as a function of SNR from this plot,
and then decreased all of our ADDSCAN 12\um\ fluxes by the amount needed to
produce average agreement with the FSC--2. We found that only very slight
changes were made in the 12\um\ LF.

In conclusion, at lower redshifts, our ADDSCAN fluxes are slightly larger than
those in the FSC-2 due to the resolution of extended emission by the FSC-2
spatial template, especially at 12 and 25\um.  At higher redshifts, the
over--estimation of Fnu(z) also contributes, but the resolution effect is still
probably a dominant contributor, since at 60 and 100\um\, where the resolution
effect is least significant, the ADDSCAN and FSC-2 fluxes become virtually
identical. Comparing the small flux over--estimate made by Fnu(z) for low--SNR
sources, to the large under--estimation of fluxes in the FSC--2 and PSC--2, we
conclude that ADDSCAN/SCANPI data, while imperfect, are the best method
available to obtain accurate {\it total galaxy} flux densities.

\subsection*{2. ADDSCAN--FSC--2 COLOR DIFFERENCES VS. REDSHIFT}

We have also plotted the difference in infrared colors obtained from the
ADDSCAN
and FSC--2 flux densities. IF beam resolution strongly affected the observed
infrared colors of the sample galaxies, we would see the effects in these
graphs.  As is evident, a best-fit line to the points shows no effect at any
color for the Seyfert galaxies and small effects for the normal galaxies.  The
slopes for the normal galaxies show the FSC--2 colors to be redder at lower
redshifts, where beam size is more important, as is expected from the large
beams at longer wavelengths.  The only exception is the 12--25\um\ color, which
compares two wavebands with the same detector size (the 25\um\ detectors, on
the
average, are larger by $\sim 3\%$, {\it IRAS Explanatory Supplement 1988}).
The
slope in this graphs goes slightly the other way (redder color at higher
redshifts), probably because the higher--luminosity objects have relatively
stronger 25\um\ emission. We note that, since the intrinsic scatter in the
colors is large compared to the possible effects of beam size, this test is not
very sensitive, and thus should only be taken as a consistency check.
\vfill\eject

\subsection*{REFERENCES}
\addtocounter{page}{+7}

\rf{Becker, R. H., White, R. L., \& Edwards, A. L. 1991, \aps{75}{1}}
\rf{Bicay, M. D. \& Helou, G. 1990, \apj{362}{59}}
\rf{Condon, J. J. 1992, \ara{30}{575}}
\rf{Condon, J. J., Condon, M., A., Gisler, G., \& Puschell, J. 1982,
\apj{252}{102}}
\rf{de Vaucouleurs, G., de Vaucouleurs, A., \& Corwin, H. G., Jr. 1976 Second
Reference Catalog of Bright Galaxies (Austin: University of Texas Press)}
\rf{Edelson, R. A., Malkan, M. A., \& Rieke, G. H. 1987, \apj{321}{233}}
\rf{Feigelson, E. D. \& Nelson, P. I. 1985, \apj{293}{192}}
\rf{Geller, M. J., \& Huchra, J. P. 1983, \aps{52}{61}}
\rf{Green, R. F., Schmidt, M., \& Liebert, J. 1987, \aps{61}{305}}
\rf{Heckman, T. M. 1980, \am{87}{152}}
\rf{Hewitt, A., \& Burbidge, G. 1991, \aps{75}{297}}
\rf{Hoffleit, D. 1982, The Bright Star Catalogue, Fourth Revised Edition (New
Haven: Yale University Observatory)}
\rf{Huchra, J. P., Geller, M. J., Clemens, C., Tokarz, S., \& Michel, A. 1992,
in preparation (ZCAT)}
\rf{IRAS Catalogs and Atlases: Explanatory Supplement 1988, ed. C. A. Beichman,
G. Neugebauer, H. J. Habing, P. E. Clegg, \& T. J. Chester (Washington, D. C.:
GPO)}
\rf{IRAS Point Source Catalog, Version 2. 1988, Joint IRAS Science Working
Group
(Washington, D. C.: GPO)}
\rf{Isobe, T. La~Valley, M., \& Feigelson, E. D. 1992, preprint}
\rf{Lawrence, A., Walker, D., Rowan--Robinson, M., Leech, K. J., \& Penston, M.
V. 1986, {\mn{219}{687}}
\rf{Lauberts, A. 1982 The ESO/Uppsala Survey of the ESO(B) Atlas (Munich:
European Southern Observatory)}
\rf{Low, F. J., Huchra, J. P., Kleinmann, S. G., \& Cutri, R. M. 1988,
\apj{327}{L41}}
\rf{Moshir, M., et al. 1991, Explanatory Supplement to the IRAS Faint Source
Survey, Version 2. (Pasadena: JPL)}
\rf{Norris, R. P., Allen, D. A., Sramek, R. A., Kesteven, M. J., \& Troup, E.
R.
1990, \apj{359}{291}}
\rf{Rice, W., Lonsdale, C. J., Soifer, B. T., Neugebauer, G., Kopan, E. L.,
Lloyd, L. A., deJong, T., \& Habing, H. J. 1988, \aps{68}{91}}
\rf{Sandage, A. 1975, in Stars and Stellar Systems, Vol. {\bf 9}, Galaxies and
the Universe, ed. A. Sandage, M. Sandage, \& J. Kristian (Chicago: University
of Chicago Press), p. 779}
\rf{Saunders, W. 1991, private communication}
\rf{Schwering, P. B. W., \& Israel, F. P. 1989, \as{75}{75}}
\rf{Schmidt, M. 1968, \apj{151}{393}}
\rf{Smithsonian Astrophysical Observatory Star Catalog (4 vols.) 1966, Pub.
4652
(Washington, D.C.: Smithsonian Institution)}
\rf{Soifer, B. T., Sanders, D. B., Madore, B. F., Neugebauer, G., Danielson, G.
E., Elias, J. H., Persson, C. J., \& Rice, W. L. 1987, \apj {320}{238}}
\rf{Soifer, B. T., Boehmer, G., Neugebauer, \& Sanders, D. B. 1989, \aj{98}{3}}
\rf{Soifer, B. T. \& Neugebauer, G.  1991, \aj{101}{354}}
\rf{Spinoglio, L., \& Malkan, M. A. 1989, \apj{342}{83}}
\rf{Spinoglio, L., Malkan, M. A., Rush, B., Carrasco, L., \& Recillas, E. 1993,
in preparation}
\rf{Strauss, M. A., Davis, M., Yahil, A., \& Huchra, J. P. 1990, \apj{361}{49}}
\rf{Strauss, M. A. 1991, private communication}
\rf{Strauss, M. A., Huchra, J. P., Davis, M., Yahil, A., Fisher, K. B., \&
Tonry, J. 1992, \aps{83}{29}}
\rf{V\`eron--Cetty, M. P., \& V\`eron, P. 1991, A Catalogue of Quasars and
Active Nuclei, 5th edition.  ESO Scientific Report No. 10 -- October 1991
(Munich: European Southern Observatory)}
\rf{Woltjer, L. 1990, in Active Galactic Nuclei, Blandford, R. D., Netzer, H.,
\& Woltjer, L., ed. T. J.--L. Courvoisier \& M. Mayor (Berlin:
Springer--Verlag), p.13}

\eject
\noindent {\bf FIGURE LEGENDS}
\\
\noindent {\bf Figure 1} --
The differential 12$\mu$m number counts versus log flux (from ADDSCANs), for
all
objects in our sample. The solid line is the best fit line (to all points with
a
12\um\ flux density $\ge\ .3~Jy$) with a slope of $-3/2$, as expected slope for
a complete sample with a uniform spatial distribution. The lower line is an
exponentially decreasing (towards lower flux) function, fit to all data points.
\\
\noindent {\bf Figure 2} --
{\it a}: The average values of V/$V_{max}$ for sample galaxies as a function of
their 12$\mu$m luminosity.  The error bars are the Poisson--statistical
fluctuations (point without error bars represent 3 or fewer objects.). The
dashed line shows the Euclidean value of 0.5 {\it b}: The average values of
V/$V_{max}$ for the different subsamples of galaxies as a function of their
12$\mu$m luminosity. The error bars are the Poisson--statistical fluctuations.
The dashed lines represent values of 0.5 for each subsample.
\\
\noindent {\bf Figure 3} --
Sky distribution, in galactic coordinates. Top: all galaxies in the 12\um\
sample; bottom: Seyfert galaxies in the 12\um\ sample.
\\
\noindent {\bf Figure 4} --
Distribution of distances for the various classes of galaxies: top: normal
galaxies compared to Seyfert~1 galaxies and LINERs; bottom: normal galaxies
compared to Seyfert~2 galaxies and starburst (high $L_{FIR}$) galaxies.
\\
\noindent {\bf Figure 5} --
The [60 -- 25] / [25 -- 12] two--color diagram for all the sample galaxies.
Both
axes are logarithmic. {\it a}: colors of normal galaxies, Seyfert~1s, and
Seyfert~2s. {\it b}: colors of normal galaxies, LINERs and starburst galaxies.
The horizontal dashed line indicates the maximum [60 -- 25] color allowed in
the
selection of the warm extragalactic objects by Low et al. (1988).
\\
\noindent {\bf Figure 6} --
The Venn diagram showing the overlaps among our sample, the BGS and the CfA
Seyfert galaxies.
\\
\noindent {\bf Figure 7} --
Plot of 6~cm vs. 60\um\ luminosity for all 471 12\um\ objects in the area of
sky
covered by the 4.85 GHz Survey (Becker et al. 1991). Large symbols represent
6~cm detections and small symbols represent upper limits.
\\
\noindent {\bf Figure 8} --
The 12\um\ luminosity functions. The differential space densities (per
magnitude) of the galaxies in our 12$\mu$m sample as function of the 12\um\
luminosity are plotted in logarithmic form. The error bars show the
Poisson--statistical counting uncertainties. {\it a}: Seyfert galaxies are
compared to non--Seyfert galaxies; {\it b}: Seyfert~1 galaxies are compared to
Seyfert~2 galaxies. Curves represent fits to a double power--law, as explained
in the text; {\it c}: Each galaxy type shown individually, including the lower
limits of the differential space densities of LINERs and the points referring
to
the luminosity--limited sample of starburst galaxies (see the discussion in the
text).  Points have been shifted horizontally for clarity.
\\
\noindent {\bf Figure 9} --
The far--infrared luminosity functions. The differential space densities (per
magnitude) of the galaxies in our 12\um\ sample as function of the far--IR
luminosity integrated over the four {\it IRAS} bands are plotted in logarithmic
form. The error bars show the Poisson--statistical counting uncertainties. {\it
a}: Seyfert galaxies are compared to non--Seyfert galaxies; {\it b}: Normal
galaxies are compared to Seyfert~1, Seyfert~2, LINERs and starburst galaxies.
As
discussed in the text, the space densities for LINERs are lower limits, while
those of starburst galaxies have to be considered with caution, because of the
luminosity--limited selection of this subsample. The points in {\it b} are
shifted horizontally for clarity.
\\
\noindent {\bf Figure 10} --
The differential 60\um\ number counts versus log flux for our entire sample.
The
solid line of slope $-3/2$ shows completeness down to a 60\um\ flux of 8.3 Jy.
\\
\noindent {\bf Figure 11} --
The 60\um\ space densities of the subsample of 235 galaxies complete at 60\um\
that we extracted from our 12\um\ sample are compared with the ones derived for
the BGS (Soifer et al. 1987), and for the deep high galactic latitude 60\um\
sample of Lawrence et al. (1986). Curves represent fits to a double power--law,
as explained in the text.
\\
\noindent {\bf Figure 12} --
The 60--12\um\ color as a function of 60\um\ Luminosity. All objects are
plotted.  The solid line represents a least--square--fit to the data, excluding
the Seyferts and LINERs.
\\
\noindent {\bf Figure 13} --
A comparison of our Seyfert galaxy luminosity functions with those of the CfA
sample (Edelson, Malkan, and Rieke 1987). {\it a}: The 60\um\ space densities
of
all Seyfert galaxies in our sample are compared to the ones derived for all CfA
Seyfert galaxies; {\it b}: Same as {\it a}, but for each type of Seyfert
individually.
\\
\noindent {\bf Figure B1} --
The ratio of ADDSCAN--to--FSC--2 flux densities are plotted as a function of
redshift for all normal galaxies in our sample.  Solid circles represent the
median value of the flux density ratio in each $500~km~s^{-1}$ bin of redshift.
Open squares and triangles represent the first and third quartile points in
each
bin, respectively.  Bins with only 1 or two points have only as many objects.
The few objects with $V>12,000~km~s^{-1}$. are not included. {\it a, b,
c,}~and~{\it d}: 12, 25, 60, and 100\um\ flux density, respectively.
\\
\noindent {\bf Figure B2} --
The same plots as Figure B1, for all of the Seyfert galaxies in our sample.
\\
\noindent {\bf Figure B3} --
The difference between infrared spectral slopes, as obtained from ADDSCAN and
FSC--2 flux densities, are shown as a function of redshift.  The solid line is
a
least--squares fit to the Seyferts (large circles) and the dashed line is a
least--squares fit to the non--Seyferts (small x's). {\it a, b, c,}~and~{\it
d}:
25--12\um, 60--25\um, 100--60\um, and 60--12\um\ slopes, respectively.}

\end{document}